\documentclass[aps,showpacs,pra]{revtex4}
\usepackage{epsfig,amssymb,amscd}
\begin{document}
\bibliographystyle{prsty}

\title{\bf Lower ground state due to counter-rotating wave interaction in trapped ion system}
\author{T. Liu$^{1}$, K.L. Wang$^{1,2}$, and M. Feng$^{3}$ \footnote[1]{Electronic address: mangfeng@wipm.ac.cn}} 
\affiliation{$^{1}$ The School of Science, Southwest University of Science and Technology, Mianyang 621010, China \\
$^{2}$ The Department of Modern Physics, University of Science and Technology of China, Hefei 230026, China \\
$^{3}$ State Key Laboratory of Magnetic Resonance and Atomic and Molecular Physics, Wuhan Institute of Physics and Mathematics, 
Chinese Academy of Sciences, Wuhan, 430071, China}
\date{\today}

\begin{abstract}

We consider a single ion confined in a trap under radiation of two traveling waves of lasers. In the strong-excitation regime 
and without the restriction of Lamb-Dicke limit, the Hamiltonian of the system is similar to a driving Jaynes-Cummings model 
without rotating wave approximation (RWA). The approach we developed enables us to present a complete eigensolutions, 
which makes it available to compare with the solutions under the RWA. We find that, the ground state in our non-RWA solution is
energically lower than the counterpart under the RWA. If we have the ion in the ground state, it is equivalent to a spin 
dependent force on the trapped ion.  Discussion is made for the difference between the solutions with and without 
the RWA, and for the relevant experimental test, as well as for the possible application in quantum information processing.

\end{abstract}
\vskip 0.1cm
\maketitle

\section {introduction}

Ultracold ions trapped as a line are considered as a promising system for quantum information processing \cite {cz}. Since the first 
quantum gate performed in the ion trap \cite {monroe1}, there have been a series of experiments with trapped ions to achieve nonclassical 
states \cite {wineland1}, simple quantum algorithm \cite {wineland2}, and quantum communication \cite {wineland3}.

There have been also a number of proposals to employ trapped ions for quantum computing, most of which work only in the weak excitation 
regime (WER), i.e., the Rabi frequency smaller than the trap frequency. While as bigger Rabi frequency would lead to faster quantum gating, 
some proposals \cite {cirac1,zheng,feng2} have aimed to achieve operations in the case of the Rabi frequency larger than 
the trap frequency, i.e., the so called strong excitation regime (SER). The difference of the WER from the SER is mathematically reflected 
in the employment of the rotating wave approximation (RWA), which averages out the fast oscillating terms in the interaction Hamiltonian. 
As the RWA is less valid with the larger Rabi 
frequency, the treatment for the SER was complicated, imcomplete \cite {feng1998}, and sometimes resorted to numerics \cite {zeng}.

In addition, the Lamb-Dicke limit strongly restricts the application of the trapped ions due to technical challenge and the slow quantum 
gating. We have noticed some ideas \cite {gr,duan} to remove the Lamb-Dicke limit in designing quantum gates, which are achieved by using some 
complicated laser pulse sequences. 

In the present work, we investigate, from another research angle, the system mentioned above in SER and in the absence of the Lamb-Dicke 
limit. The 
main idea, based on an analytical approach we have developed, is to check the eigenvectors and the eigenenergies of such a system, with 
which we hope to obtain new insight into the system for more application. The main result in our work is a newly found ground state, 
energically lower than the ground state calculated by standard Jaynes-Cummings model. We will also present the analytical forms of the 
eigenvectors and the variance of the eigenenergies with respect to the parameters of the system, which might be used in understanding the 
time evolution of the system.

The paper is organized as follows. In Section II we will solve the system in the absence of the RWA. Then some numerical results will 
be presented in comparison with the RWA solutions in Section III. We will discuss about the new results for their possible application.
More extensive discussion and the conclusion are made in Section IV. Some analytical deduction details could be found in Appendix.

\section {The analytical solution of the system}

As shown in Fig. 1, we consider a Raman $\Lambda$-type configuration, which corresponds to the actual process in NIST experiments. 
Like in \cite {feng1}, we will employ some unitary transformations to get rid of the assumption of Lamb-Dicke limit and the WER. So 
our solution is more general than most of the previous work \cite {previous}. For a single trapped ion experiencing two off-resonant 
counter-propagating 
traveling wave lasers with frequencies $\omega_{1}$ and $\omega_{2}$, respectively, and in the case of a large detuning $\delta$, we have 
an effective two-level system with the lasers driving the electric-dipole forbidden transition $|g\rangle$ $\leftrightarrow$ $|e\rangle$ 
by the effective laser frequency $\omega_{L}=\omega_{1}-\omega_{2}$. So we have the dimensionless Hamiltonian    
\begin {equation}
H= \frac {\Delta}{2} \sigma_{z} + a^{\dagger}a + \frac {\Omega}{2} (\sigma_{+}e^{i\eta \hat{x}} + \sigma_{-}e^{-i\eta \hat{x}}),
\end {equation}
in the frame rotating with $\omega_{L}$, where $\Delta=(\omega_{0}-\omega_{L})/\nu$, $\omega_{0}$ and $\nu$ are the resonant frequency 
of the two levels of the ion and the trap frequency, respectively. $\Omega$ is the dimensionless Rabi frequency in units of $\nu$ and 
$\eta$ the Lamb-Dicke 
parameter. $\sigma_{\pm,z}$ are usual Pauli operators, and we have $\hat{x}=a^{\dagger}+a$ for the dimensionless position operator of the ion 
with $a^{\dagger}$ and $a$ being operators of creation and annihilation of the phonon field, respectively. We suppose that both $\Omega$ and 
$\nu$ are much larger than the atomic decay rate and the phonon dissipative rate so that no dissipation is considered below.

Like in \cite {feng1}, we first carry out some unitary transformations on Eq. (1) to avoid the expansion of the exponentials. So we have
\begin {equation}
H^{I} = UHU^{\dagger} = \frac {\Omega}{2} \sigma_{z} + a^{\dagger}a + g (a^{\dagger} + a) \sigma_{x} + \epsilon \sigma_{x} + g^{2},
\end {equation}
where
$$U=\frac {1}{\sqrt{2}}e^{i\pi a^{\dagger}a/2} \pmatrix {F^{\dagger}(\eta) & F(\eta) \cr -F^{\dagger}(\eta) & F(\eta)},$$
with $F(\eta)=\exp {[i\eta (a^{\dagger} + a)/2]}$, $g=\eta/2$, and $\epsilon=-\Delta/2$. Eq. (2) is a typical driving Jaynes-Cummings 
model including the counter-rotating wave terms. In contrast to the usual treatments to consider the Lamb-Dicke limit by using the RWA 
in a frame rotation, we remain the counter-rotating wave interaction in the third term of the right-hand side of Eq. (2) in our case. 
To go on our treatment, we make a further rotation with $V=\exp{(i\pi\sigma_{y}/4)}$, yielding
\begin {equation}
H^{'} = VH^{I}V^{\dagger} = -\frac {\Omega}{2} \sigma_{x} + a^{\dagger}a + g (a^{\dagger} + a) \sigma_{z} + \epsilon \sigma_{z} + g^{2},
\end {equation}
where we have used $\exp {(i\theta\sigma_{y})}\sigma_{x}\exp {(-i\theta\sigma_{y})}=\cos(2\theta)\sigma_{x} + \sin (2\theta)\sigma_{z}$,
and $\exp {(i\theta\sigma_{y})}\sigma_{z}\exp {(-i\theta\sigma_{y})}=\cos(2\theta)\sigma_{z} - \sin (2\theta)\sigma_{x}$. For convenience of 
our following treatment, we rewrite Eq. (3) to be
\begin {equation}
H^{'} = \epsilon (|e\rangle\langle e| - |g\rangle\langle g|) - \frac {\Omega}{2} (|e\rangle\langle g| + |g\rangle\langle e|) + 
a^{\dagger}a + g (a^{\dagger} + a) (|e\rangle\langle e| - |g\rangle\langle g|) + g^{2}.
\end {equation}

Using Schr\"odinger equation, and the orthogonality between $|e\rangle$ and $|g\rangle$, we suppose
\begin {equation}
|\rangle= |\varphi_{1}\rangle|e\rangle + |\varphi_{2}\rangle|g\rangle,
\end {equation}
which yields
\begin {equation}
\epsilon |\varphi_{1}\rangle + a^{\dagger}a |\varphi_{1}\rangle + g(a^{\dagger} + a) |\varphi_{1}\rangle - 
\frac {\Omega}{2}|\varphi_{2}\rangle + g^{2}|\varphi_{1}\rangle = E |\varphi_{1}\rangle,
\end {equation}
\begin {equation}
-\epsilon |\varphi_{2}\rangle + a^{\dagger}a |\varphi_{2}\rangle - g(a^{\dagger} + a) |\varphi_{2}\rangle - 
\frac {\Omega}{2}|\varphi_{1}\rangle + g^{2}|\varphi_{2}\rangle = E |\varphi_{2}\rangle.
\end {equation}
To make the above equations concise, we apply the displacement operator $\hat{D}(g)=\exp{[g(a^{\dagger}-a)]}$ on $a^{\dagger}$ and $a$, 
which gives $A=\hat {D}(g)^{\dagger} a \hat {D}(g)= a+g$, $A^{\dagger}= \hat {D}(g)^{\dagger} a^{\dagger} \hat {D}(g) = a^{\dagger} + g$, 
$B= \hat {D}(-g)^{\dagger} a \hat {D}(-g) = a-g$, and $B^{\dagger}= \hat {D}(-g)^{\dagger} a^{\dagger} \hat {D}(-g) = a^{\dagger} - g$. 
So we have
\begin {equation}
(A^{\dagger}A + \epsilon) |\varphi_{1}\rangle - \frac {\Omega}{2} |\varphi_{2}\rangle = E |\varphi_{1}\rangle,
\end {equation}
\begin {equation}
(B^{\dagger}B - \epsilon) |\varphi_{2}\rangle - \frac {\Omega}{2} |\varphi_{1}\rangle = E |\varphi_{2}\rangle.
\end {equation}
Obvious, the new operators work in different subspaces, which leads to different evolutions regarding different internal levels $|g\rangle$
and $|e\rangle$. We will later refer to this feature to be relevant to spin-dependent force. The solution of the two equations above can 
be simply set as
\begin{equation}
|\varphi_{1}\rangle = \sum_{n=0}^{N} c_{n} |n\rangle_{A},  
\end{equation}
\begin{equation}
|\varphi_{2}\rangle = \sum_{n=0}^{N} d_{n} |n\rangle_{B},  
\end{equation}
with N a large integer to be determined later, $|n\rangle_{A}=\frac {1}{\sqrt{n!}}(a^{\dagger}+g)^{n} |0\rangle_{A} = \frac {1}{\sqrt{n!}}
(a^{\dagger}+g)^{n} \hat {D}(g)^{\dagger}|0\rangle =\frac {1}{\sqrt{n!}}(a^{\dagger}+g)^{n}\exp\{-ga^{\dagger}-g^{2}/2\}|0\rangle,$
and $|n\rangle_{B}=\frac {1}{\sqrt{n!}}(a^{\dagger}-g)^{n} |0\rangle_{B} = \frac {1}{\sqrt{n!}}(a^{\dagger}-g)^{n} \hat {D}(-g)^{\dagger} 
|0\rangle=\frac {1}{\sqrt{n!}}(a^{\dagger}-g)^{n}\exp\{ga^{\dagger}-g^{2}/2\}|0\rangle$. Taking Eqs. (10) and (11) into Eqs. (8) and (9), 
respectively, and multiplying by $_{A}\langle m|$ and $_{B}\langle m|$, respectively, we have,
\begin {equation}
(m+\epsilon) c_{m} - \frac {\Omega}{2} \sum_{n=0}^{N} (-1)^{n} D_{mn} d_{n} = Ec_{m},
\end {equation}
\begin {equation}
(m-\epsilon) d_{m} - \frac {\Omega}{2} \sum_{n=0}^{N} (-1)^{m} D_{mn} c_{n} = Ed_{m},
\end {equation}
where we have set $(-1)^{n}D_{mn}= _{A}\langle m|n\rangle_{B}$ and $(-1)^{m}D_{mn} = _{B}\langle m|n\rangle_{A}$, whose deduction 
can be found in Appendix. Diagonizing the relevant determinants, we may have the eigenenergies $E_{i}$ and the 
eigenvectors regarding $c_{n}^{i}$ and $d_{n}^{i}$ ($n=0, \cdots, N, i= 0, \cdots, N$). Therefore, as long as we could find a closed subspace
with $c_{N+1}^{i}$ and $d_{N+1}^{i}$ approaching zero for a certain big integer N, we may have a complete eigensolution of the system.

\section {discussion based on numerics}

Before doing numerics, we first consider a treatment by involving the RWA. As the RWA solution could present complete eigenenergy 
spectra, it is interesting to make a comparison between the RWA solution and our non-RWA one. We consider a rotation in Eq. (2) 
with respect to $\exp \{-i[(\Omega/2)\sigma_{z} + a^{\dagger}a]t\}$, which results in
\begin {equation}
H_{A}=\frac {\Omega}{2}\sigma_{z} + a^{\dagger}a + g (a\sigma_{+} + a^{\dagger}\sigma_{-}) + g^{2},
\end {equation}
where the RWA has been made by setting $\Omega=1$, and we have corresponding eigenenergies 
\begin {equation}
E^{\pm}_{n}=(n+g^{2}+1/2)\pm g\sqrt{n+1}.
\end {equation}
So the system is degenerate in the case of $\eta=0$ and there are two eigenenergy spectra corresponding to $E^{\pm}_{n}$ as 
long as $\eta\neq 0$.

Figs. 2(a) and 2(b) demonstrate two spectra, respectively, and in each figure we compare the differences between the RWA and non-RWA 
solutions \cite {explain2}. In contrast to the two spectra in the RWA solution, 
the non-RWA solution includes only one spectrum. Comparing the two eigensolutions, we find that the
even-number and odd-number excited levels in the non-RWA case correspond to $E^{+}_{n}$ and $E^{-}_{n}$ of the RWA case, respectively, and 
the difference becomes bigger and bigger with the increase of $\eta$. It is physically understandable for these differences because the 
RWA solution, valid only for small $\eta$, does not work beyond the Lamb-Dicke regime. Above comparison also demonstrates the change of 
the ion 
trap system from an integrable case (i.e., with RWA validity) to the non-integrable case (i.e., without RWA validity). But besides these 
differences, we find an unusual result in this comparison, i.e., a new level without the counterpart in RWA solution appearing in our 
solution, which is lower than the ground state in RWA solution by 
$\nu + x\eta$ with $x$ a $\eta$-dependent coefficient. In the viewpoint of physics, due to additional counter-rotating wave interaction
involved, it is reasonable to have something more in our solution than the RWA case, although this does not surely lead to a new level lower 
than the previous ground state. Anyway, this is a good news for quantum information processing with trapped ions. As the situation in SER 
and beyond the Lamb-Dicke limit involves more instability, a stable confinement of the ion requires a stronger trapping condition. In this 
sense, our solution, with the possibility to have the ion stay in an energically lower state, gives a hope in this respect. 
We will come to this point again later.

Since no report of the new ground state had been found either theoretically or experimentally in previous publications,
we suggest to check it experimentally by resonant absorption spectrum. As shown above, in the case of non-zero Lamb-Dicke
parameter, the degeneracy of the neighboring level spacing is released, and the bigger the $\eta$, the larger the spacing
difference between the neighboring levels. Therefore, an experimental test of the newly found ground state should be
available by resonant transition between the ground and the first excited states in Fig. 2, once the SER is reached.
We have noticed that the SER could be achieved by first cooling the ions within the Lamb-Dicke limit and under the
WER, and then by decreasing the trap frequency by opening the trap adiabatically \cite {cirac1}.

Since it is lower in energy than the previously recognized ground states, the new ground state we found is more
stable, and thereby more suitable to store quantum information. Once the trapped ion is cooled down to the ground state in the SER, 
it is, as shown in Eq. (5) with $n=0$, actually equivalent to the effect of a  spin-dependent force on the trapped ion \cite {hal}. If 
we make Hadamard gate on the ion by $|g\rangle\rightarrow (|g\rangle + |e\rangle)/\sqrt{2}$
and $|e\rangle\rightarrow (|g\rangle - |e\rangle)/\sqrt{2}$, we reach a Schr\"odinger cat state, i.e., 
$(1/2)\{[ D^{\dagger}(g)|0\rangle + D^{\dagger}(-g)|0\rangle]|g\rangle - [D^{\dagger}(g)|0\rangle - D^{\dagger}(-g)|0\rangle]|e\rangle\}$.
Two ions confined in a trap in above situation will yield two-qubit gates without really exciting the vibrational mode \cite {gr}. It
is also the way with this spin-dependent force towards scalable quantum information processing \cite {duan}. As in SER, we may have larger 
Rabi frequency than in WER, the quantum gate could be in principle carried out faster in the SER.

In addition, as it is convergent throughout the parameter subspace, our complete eigensolution enables us to 
accurately write down the state of the system at an arbitrary evolution time,
provided that we have known the initial state. This would be useful for future experiments in preparing non-classical states
and in designing any desired quantum gates with trapped ions in the SER and beyond the Lamb-Dicke limit.  Moreover,
as shown in Figs 3(a), 3(b) and 3(c), our present solution is helpful for us to understand the particular solutions in 
previous publication \cite {feng1}. The
comparison in the figures shows that the results in \cite {feng1} are actually mixtures of different eigensolutions. For example,
the lowest level in Fig. 2 in \cite {feng1}, corresponding to $\Omega = 2$ and $\eta = 0.2$, is actually constituted at least by 
the third, the fourth, and the fifth excited states of the eigensolution. 

\section {further discussion and conclusion}

The observation of the counter-rotating effects is an interesting topic discussed previously. In \cite {crisp}, a standard method is 
used to study the observable effects regarding the rotating and the counter-rotating terms in the Jaynes-Cummings model, including to 
observe Bloch-Siegert shift \cite {bloch} and quantum chaos in a cavity QED by using differently polarized lights. A recent work \cite {jan} 
for a two-photon Jaynes-Cummings model has also 
investigated the observability of the counter-rotating terms. By using perturbation theory, the authors claimed that the counter-rotating
effects, although very small, can be in principle observed by measuring the energy of the atom going through the cavity. 
Actually, for the cavity QED system without any external source involved, it 
is generally thought that the counter-rotating terms only make contribution in some virtual fluctuations of the energy
in the weak coupling regime. While the interference between the rotating and counter-rotating contributions could result in some 
phase dependent effects \cite {phoenix}. Anyway, if there is an external source, for example, the laser radiating a trapped ultracold ion, 
the counter-rotating terms will show their effects, e.g., related to heating in the case of WER \cite {single}. In this sense, our 
result is somewhat amazing because the counter-rotating interaction in the SER, making entanglement between internal and vibrational states 
of the trapped ion, plays positive role in the ion trapping.

We argue that our approach is applicable to different physical processes involving counter-rotating interaction.
Since the counter-rotating terms result in energy nonconservation in single quanta processes, usual techniques cannot solve the 
Hamiltonian with eigenstates spanning in an open form. In this case, path-integral approach \cite {zz} and perturbation approach 
\cite {phoenix}, assisted by numerical techniques were employed in the weak coupling regime of the Jaynes-Cummings model. In contrast, our
method, based on the diagonalization of the coherent-state subspace, could in principle  study the Jaynes-Cummings model without the RWA in 
any cases. We have also noticed a recent publication \cite {irish} to treat a strongly coupled two-level system
to a quntum oscillator under an adiabatic approximation, in which something is similar to our work in the solution of the Hamiltonian in 
the absence of the RWA. But due to the different features in their system from our atomic case, the two-level splitting term, much smaller
compared to other terms, can be taken as a perturbation. So the advantage of that treatment is the possibility to analytically obtain good
approximate solutions. In contrast, not any approximation is used in our solution, which should be more efficient
to do the relevant job.

In summary, we have investigated the eigensolution of the system with a single trapped ion, experiencing two traveling waves of lasers, 
in the SER and in the absence of the Lamb-Dicke limit. We have found the ground state in the non-RWA case to be 
energically lower than the counterpart of the solution with RWA, which would be useful for quantum information storage and for quantum 
computing. The analytical forms of the eigenfunction and the complete set of the eigensolutions would 
be helpful for us to understand a trapped ion in the SER and with a large Lamb-Dicke parameter. We argue that our work would be applied 
to different systems in dealing with strong coupling problems.

\section {acknowledgments}

This work is supported in part by NNSFC No. 10474118, by Hubei Provincial Funding for Distinguished Young Scholars, 
and by Sichuan Provincial Funding.

\section {appendix}

We give the deduction of $_{A}\langle m|n\rangle_{B}$ and  $_{B}\langle m|n\rangle_{A}$  below,
$$_{A}\langle m | n \rangle_{B} = \frac {1}{\sqrt{m!n!}} \langle 0|e^{-ga-g^{2}/2}(a+g)^{m}(a^{\dagger}-g)^{n}
e^{ga^{\dagger}-g^{2}/2}|0\rangle $$
$$ = \frac {1}{\sqrt{m!n!}} e^{-2g^{2}}\langle 0|(a+g)^{m}e^{ga^{\dagger}}e^{-ga}(a^{\dagger}-g)^{n} |0 \rangle$$
$$ = \frac {1}{\sqrt{m!n!}} e^{-2g^{2}}\langle 0|(a+2g)^{m}(a^{\dagger}-2g)^{n}|0 \rangle = (-1)^{n} D_{mn},$$
with 
$$D_{mn}=e^{-2g^{2}}\sum_{i=0}^{min[m,n]} (-1)^{-i}\frac {\sqrt{m!n!}(2g)^{m+n-2i}}{(m-i)!(n-i)!i!}.$$

It is easily proven following a similar step to above that   
$$_{B}\langle m | n \rangle_{A} = \frac {1}{\sqrt{m!n!}} \langle 0|e^{ga-g^{2}/2}(a-g)^{m}(a^{\dagger}+g)^{n}
e^{-ga^{\dagger}-g^{2}/2}|0\rangle, $$
would finally get to $(-1)^{m}D_{mn}.$

\begin {center}{\bf The captions of the figures}\end{center}

Fig. 1 Schematic of a single trapped ion under radiation of two traveling wave lasers, where $\omega_{1}$ and $\omega_{2}$ are frequencies 
regarding the two lasers, respectively, $\omega_{0}$ is the resonant frequency between $|g\rangle$ and $|e\rangle$, and $\delta$ and 
$\Delta$ are relevant detunings. This is a typical Raman process employed in NIST experiments, with for example $Be^{+}$, for quantum 
computing. 

Fig. 2 The eigenenergy spectra with $\Omega=1$, where (a) and (b) correspond to two different sets of eigenenergies with respect to
Lamb-Dicke parameter. In (a) the comparison is made between $E^{+}_{n}$ in the RWA case (dashed-dotted curves) and $E_{n}$ with $n=$ even 
numbers in the non-RWA case (star curves for $n=0$ and solid curves for others);
In (b) the comparison is for $E^{-}_{n}$ in the RWA case (dashed-dotted curves) to $E_{n}$ with $n=$ odd numbers in the non-RWA case 
(solid curves).

Fig. 3 The eigenenergy with respect to the detuning $\Delta$, where for convenience of comparison we have used the same parameter 
numbers as in \cite {feng1}. For clarity, we plot the different levels with different lines. 
The parameter numbers are $\Omega=2$, and (a) $\eta=0.2$; (b) $\eta=0.4$; (c) $\eta=0.6$. 


\begin{thebibliography}{22}
\bibitem{previous} Most of the previous work in this respect were carried out by cuting off the expansion of the exponentials regarding 
the quantized phonon operators, which is only reasonable in the WER and within the Lamb-Dicke limit. In contrast, our treatment can be used
in both the SER and the WER cases.
\bibitem {explain2} We take throughout this paper $N=40$ in which the coefficients $c_{41}^{i}$ and $d_{41}^{i}$ 
with $i=0, 1, .. 40 $ are negligible in the case of $\Omega=1$ and 2. Although with the 
increase of values of $\Omega$ the diagonalization space has to be enlarged, our 
analytical method generally works well in a wide range of parameters.
\end{thebibliography}
\end{document}